\newcommand{\version}{May 9, 2015}
\documentclass[a4paper,twoside,11pt,pdftex]{article}
\usepackage[bindingoffset=0.2cm,textheight=22.6cm,hdivide={2.7cm,*,2.7cm}, vdivide={*,22cm,*}]{geometry}
\usepackage{amsmath,amsfonts,amssymb}
\usepackage[pdftex]{graphicx}
\usepackage[utf8]{inputenc}
\usepackage{tensind}
\tensordelimiter{?}
\IfFileExists{dsfont.sty}
	{\usepackage{dsfont}
         \let\mathbb=\mathds
         \newcommand{\id}{\mathds{1}}}
	{\typeout{Package dsfont.sty was not found, using alternative macros.}
         \let\mathds=\mathbb
         \newcommand{\id}{\mbox{1 \kern-.59em \textrm{l}}}}

\usepackage[pdftex,bookmarksnumbered=true,breaklinks=true]{hyperref}
\usepackage[numbers,square,comma,sort&compress]{natbib}

\newcommand{\EMT}{EMT }

\newcommand{\mg}{\textbf{\texttt{g}}}

\newcommand{\uim}{UV/IR mixing}
\newcommand{\nc}{non-commu\-ta\-tive}

\newcommand{\tr}{\textrm{tr}}

\newcommand{\eqnref}[1]{Eqn.~(\ref{#1})}		
\newcommand{\secref}[1]{Section~\ref{#1}}		
\newcommand{\starco}[2]{\left[ #1\stackrel{\star}{,}#2\right] }		
\newcommand{\staraco}[2]{\left\{ #1\stackrel{\star}{,}#2\right\} }	
\newcommand{\var}[2]{\frac{\d #1}{\d #2}}				
\newcommand{\pa}{\partial}						
\newcommand{\ri}{\textrm{i}}						
\newcommand{\re}{\textrm{e}}						

\renewcommand{\a}{\alpha}
\renewcommand{\b}{\beta}
\newcommand{\g}{\gamma}
\renewcommand{\d}{\delta}
\newcommand{\e}{\epsilon}

\renewcommand{\th}{\theta}

\renewcommand{\l}{\lambda}
\newcommand{\m}{\mu}
\newcommand{\n}{\nu}

\renewcommand{\r}{\rho}
\newcommand{\s}{\sigma}

\newcommand{\inv}[1]{\frac{1}{#1}}				

\newcommand{\intx}{\int\! d^4x}						
\newcommand{\nn}{\nonumber}

\newcommand{\ig}{\textrm{i}g}

\newcommand{\bpsi}{\bar{\psi}}

\newcommand{\bD}{{\bar D}}
\newcommand{\bT}{\mathbf{T}}

\title{\begin{flushright}
        {\small LA-UR-15-21030}
       \end{flushright}\vspace{2em}
       On the energy-momentum tensor in Moyal space}
\date{\version}

\author{Herbert Balasin\footnotemark[1]~, Daniel N. Blaschke\footnotemark[2]~, Fran\c{c}ois Gieres\footnotemark[3]~ \\and Manfred Schweda\footnotemark[1]}

\begin{document}

\maketitle
\thispagestyle{empty}
\begin{center}
\renewcommand{\thefootnote}{\fnsymbol{footnote}}
\vspace{-0.3cm}\footnotemark[1]Institute for Theoretical Physics, Vienna University of Technology\\Wiedner Hauptstra\ss e 8-10, A-1040 Vienna (Austria)\\[0.3cm]
\footnotemark[2]Los Alamos National Laboratory, Theoretical Division\\Los Alamos, NM, 87545, USA\\[0.3cm]
\footnotemark[3]Universit\'e de Lyon, Universit\'e Claude Bernard Lyon 1 and CNRS/IN2P3,\\Institut de Physique Nucl\'eaire, Bat. P. Dirac,\\4 rue Enrico Fermi, F-69622-Villeurbanne (France)
\\[0.5cm]
\ttfamily{E-mail: hbalasin@tph.tuwien.ac.at, dblaschke@lanl.gov, gieres@ipnl.in2p3.fr, mschweda@tph.tuwien.ac.at}
\end{center}

\vspace{1.5em}
\begin{abstract}
We study the properties of the energy-mo\-mentum tensor of gauge fields coupled to matter in non-commutative (Moyal) space.
In general, the non-commutativity
affects the usual conservation law of the tensor as well
as its transformation properties (gauge covariance instead of gauge invariance).
It is known that the conservation of the energy-momentum tensor can be achieved by a redefinition involving another star-product.
Furthermore, for a pure gauge theory  it is always possible to define a gauge invariant energy-momentum tensor
by means of a Wilson line.

We show that the latter two procedures are incompatible with each other if couplings of gauge fields to matter fields (scalars or fermions) are considered:
The gauge invariant tensor (constructed via Wilson line) does not allow for a redefinition
assuring its conservation, and
vice-versa the introduction of another star-product does not allow for gauge invariance by means of a Wilson line.
\end{abstract}

\newpage
\tableofcontents

%

\section{Introduction}

Groenewold-Moyal  (or $\th$-deformed) space~\cite{Groenewold:1946,Moyal:1949}
represents one of the simplest models for quantized spaces, and has been
extensively studied over the past twenty years,
e.g. see~\cite{Szabo:2001,Rivasseau:2007a,Blaschke:2010kw} and references therein for a review.
The theories on this space are
formulated in terms of ordinary functions by means of a deformed product,
the so-called star-product
\begin{align}
(f\star g) (x) = \re^{\frac{\ri}2 \th^{\mu\nu} \partial^x_\mu \partial^y_\nu } f(x) g(y) \Big|_{y= x}
\,,
\end{align}
which implies that the coordinates fulfill
\begin{align}
\starco{x_\mu}{x_\nu} \equiv x_{\mu} \star x_{\nu} - x_{\nu} \star x_{\mu} = \ri\th_{\mu\nu}
\,.
\end{align}
This commutation relation is invariant under translations of the space-time coordinates and under the so-called reduced Lorentz transformations (or reduced orthogonal transformations in the Euclidean setting), see for instance reference~\cite{Grosse:2011es}.
In general, field theoretic models on such spaces suffer from a new  type of divergences which arise due to a phenomenon referred to as {\uim}~\cite{Minwalla:1999px,Matusis:2000jf}
and which render the models non-renormalizable.
This problem can be overcome in the case of some special scalar field models~\cite{Grosse:2003, Grosse:2004b,Gurau:2009},
one of them having been shown to be
solvable even non-perturbatively~\cite{Grosse:2012uv}.

The present work is devoted to a basic aspect of classical field theories on Moyal space, namely the
\emph{energy-momentum tensor} (hereafter referred to as \emph{EMT}) and its properties at tree level.
In earlier studies~\cite{Micu:2000,Pengpan:2000kd,Schweda:2000,AbouZeid:2001up,Grimstrup:2002xs,Das:2002jd}
some modifications to the conservation law of the \EMT
due to the non-commutativity parameters $\theta^{\mu \nu}$  were found in $\phi^{\star4}$ and in gauge theory (without matter couplings).
Here, we wish to investigate more generally complex scalars and fermions coupled to
$U_{\star}(1)$ gauge fields.
In view of the infamous time-ordering problems in quantum field theory on Moyal space~\cite{Bahns:2002},
we restrict ourselves to the Euclidean version of Moyal space.

The present work is organized as follows.
In \secref{sec:gaugefields}, we examine the gauge invariance and conservation properties of the \EMT for a gauge field in Moyal space.
In Sections \ref{sec:matterfields1} and \ref{sec:matterfields2} we then extend the discussion to include various couplings to matter.

\section{\EMT for a gauge field in Moyal space}
\label{sec:gaugefields}

We consider a $U_\star(1)$ gauge field $(A^{\mu})$ coupled to an external current $(J^\m)$ in
four-dimensional flat Euclidean\footnote{We recall that certain signs (e.g. some global signs in the actions) change upon passage from
Minkowskian to Euclidean signature.
The coupling constant is denoted by $g$ and there should be no risk of confusion
with the determinant of the metric tensor $(g_{\m \n})$ considered
for defining the Einstein-Hilbert EMT.}
Moyal space: the action
\begin{align}
 S[A]&=\inv{4}\intx\, F_{\m\n}\star F^{\m\n}+\intx\, J^\m\star A_\m
 \,,
  \label{eq:gaugeaction}
 \\
 \textrm{with} \ \; F_{\m\n}&=\pa_\m A_\n-\pa_\n A_\m-\ig\starco{A_\m}{A_\n}
 \,,
 \nonumber
\end{align}
yields the equation of motion
\begin{align}
0=  \var{S[A]}{A_\n}=-D_\m F^{\m\n}+J^\n
 \,. \label{eq:A-eom}
\end{align}
The functional  $S[A]$  is invariant under the infinitesimal gauge transformations
\begin{align}
 \d_\l A_\m&=D_\m\l \equiv
 \pa_\m\l-\ig\starco{A_\m}{\l}\,, &
 \d_\l F_{\m\n}&=-\ig\starco{F_{\m\n}}{\l}
 \label{eq:gftrafo}
 \,,
\end{align}
provided the current $(J^\m)$ does not transform and is covariantly conserved, i.e. $D_\m J^\m=0$.
This is also consistent with the equation of motion in the sense that
\begin{align}
D_\n J^\n = D_\n (D_\m F^{\m\n})
= \frac{1}{2} \,  \starco{D_{\n}}{D_{\m}} F^{\m\n}
= \frac{\ig}{2}\starco{F_{\m\n}}{F^{\m\n}}=0
 \,.
 \label{eq:covconserv}
\end{align}
Notice, however, that gauge invariance of the equation of motion \eqref{eq:A-eom} requires that $J^\m$ transforms covariantly,
i.e. $ \d_\l J^\n = - \ig \starco{J^\n}{\l}$,
but that would destroy gauge invariance of the action unless $\pa_\m J^\m=0$.
This inconsistency was already noticed in Ref.~\cite{Adorno:2011wj} and is due to the non-Abelian nature of {\nc} gauge theory.
In fact, a similar inconsistency
occurs in Yang-Mills theory on ordinary commutative space
when coupling the gauge field to an external current~\cite{Sikivie:1978sa}.
This problem can be overcome by coupling the gauge field to dynamical complex scalar and/or fermion fields
so that the external current is replaced by the corresponding matter current, see  next sections.
For now, we keep in mind that the action
\eqref{eq:gaugeaction}  is not  the complete action.

Concerning the transformation laws~\eqref{eq:gftrafo} we
emphasize that, by contrast to a $U(1)$ gauge theory
in ordinary Minkowski space, the field strength $F_{\m \n}$ is not a gauge invariant quantity
as in electrodynamics. This non-Abelian nature of the theory in Moyal space is due to the non-commutativity
of space-time coordinates
which implies that the field strength
``feels'' the non-commutativity of the space in which it lives.
(This even applies to the simplest case of a constant field strength~\cite{Balasin:2014dma}.)
The transformation law of $F_{\m \n}$ implies that
the Lagrangian density ${\cal L} = \inv{4} \, F_{\m\n}\star F^{\m\n}$
is \emph{not} invariant under gauge transformations since $\d_\l {\cal L} = -\ig\starco{{\cal L}}{\l}$:
it is only the integral which plays the role of a trace which ensures cyclic invariance of factors and thereby
gauge invariance. Henceforth, the lack of gauge invariance of the \EMT will
 not come as a surprise and contrasts the situation
for non-Abelian Yang-Mills fields in Minkowski space.

The improved \EMT for a free (i.e. not coupling to a current) gauge field in Moyal space was already computed in Ref.~\cite{AbouZeid:2001up,Grimstrup:2002xs,Das:2002jd}:
\begin{align}
 T^{\m\n}= \inv2\left(\staraco{F^{\m\r}}{F^{\n}_{\ \r}}-\inv2
 \delta^{\m \n}
 F_{\r\s}\star F^{\r\s}\right)
 \,.
 \label{eq:EMTNCgauge}
\end{align}
It is symmetric and traceless, and it transforms covariantly under gauge transformations:
\begin{align}
 \d_\l T^{\m\n}&=-\ig\starco{T^{\m\n}}{\l}
 \,.
 \label{eq:covtranslaw}
\end{align}
From the Bianchi identity
$D_\m F_{\n\r}+D_\n F_{\r\m}+D_\r F_{\m\n}=0$
and the equation of motion \eqref{eq:A-eom}
with $J^\m=0$,  it follows that the covariant divergence of the gauge field \EMT vanishes,
\begin{align}
 D_\m T^{\m\n} &= 0
\,, \label{eq:covconslaw}
\end{align}
i.e. $T^{\m\n}$ is covariantly conserved.

In the Minkowskian version of Moyal space with non-commutativity  parameters satisfying $\th^{0i}=0$,
the integral $\int d^3x$ of a star-commutator vanishes (assuming as usual that fields vanish sufficiently fast at spatial infinity),
hence equation~\eqref{eq:covconslaw} implies that
\begin{align}
0 = \int d^3x \,  D_\m T^{\m\n} &= \int d^3x \;  \pa_0 T^{0\n} = \frac{d P^\n}{dt} \, , \qquad {\rm with} \ \;
P^\n \equiv \int d^3x \,  T^{0\n}
\,.
\label{eq:intlcl}
\end{align}
Thus, the four-momentum $(P^\n )$ of the gauge field represents a conserved quantity.
Moreover, this quantity is gauge invariant by virtue of~\eqref{eq:covtranslaw}
and the definition of $P^\nu$ in terms of $T^{0\nu}$.

Let us now come back to the local transformation law~\eqref{eq:covtranslaw}.
In Ref.~\cite{Gross:2000ba} (see also~\cite{Berenstein:2001dw}) it was explained how to construct gauge invariant objects
in Moyal space  out of gauge covariant ones. In fact,
this task is achieved by folding the quantity in question with a straight Wilson line defined by a length vector $(l^\m )$ with
$l^\m=\th^{\m\n}k_\n \equiv (\theta k)^\m
$.
Using this procedure, the authors of reference~\cite{AbouZeid:2001up}
obtained a standard local conservation law for the so constructed EMT.
In the following we will also
follow this strategy for gauge fields, scalars and fermions,
and therefore we briefly outline the procedure here.

The {\nc} generalization of a straight Wilson line with the appropriate length
is given by
\begin{align}
 W(k,x)={\cal P}_\star\exp\left(\int_0^1\!d\s\,A_\m(x+\s\th k) \, \th^{\m\n}k_\n\right)
 \label{eq:NCWL}
 \,,
\end{align}
where ${\cal P}_\star$ denotes path ordering with respect to the contour parameter $\s$.
The expression \eqref{eq:NCWL} transforms as $W(k,x)\to U(x)\star W(k,x)\star U(x+\th k)^\dagger$ under a gauge transformation $U(x)$.
Hence, $\intx W(k,x)\star\exp(\ri kx)$ is a gauge invariant object because the length vector of the Wilson line
is adjusted to be $\th^{\m\n}k_\n$ and $\exp(\ri kx)$ induces a translation of $U^\dagger$ by $-\th k$, cf.~\cite{Gross:2000ba,Ishibashi:1999hs}.
One may now construct (Fourier transforms of) gauge invariant objects from gauge covariant ones by star-multiplication with $W(k,x)$ and $\exp(\ri kx)$ and integrating over $d^4x$.
The choice of a straight Wilson line is the most natural one because for such a line
it makes no difference if the operator is attached to an endpoint of the Wilson line or somewhere in the middle~\cite{Gross:2000ba}.
Furthermore, in the commutative limit $(\th\to0 )$ the Wilson line's length goes to zero.

For the \EMT of a gauge field in Moyal space this means that
\begin{align}
 \tilde{T}^{\m\n}(y)& \equiv
 \int\!\frac{d^4k\, d^4x}{(2\pi)^4} \, \re ^{\ri k(y-x)}\star W(k,x)\star T^{\m\n}(x)
 \label{eq:EMTGFM}
\end{align}
is a gauge invariant quantity\footnote{When considering $U_\star(N)$ gauge fields rather than $U_\star(1)$
fields as we do here,
an additional trace appears in the product, i.e. $W\star T$ becomes $\tr \, (W\star T)$.}
(which reduces in the commutative limit
to the ordinary \EMT due to $\lim\limits_{\th\to0} W(k,x)=1$).
However, it is not conserved~\cite{AbouZeid:2001up},
\begin{align}
 \pa^y_\m\tilde{T}^{\m\n}(y)&=\int\!\frac{d^4k \, d^4x}{(2\pi)^4} \, \re ^{\ri k(y-x)}\star \left(\pa^x_\m W(k,x)\star T^{\m\n}(x)+W(k,x)\star \pa^x_\m T^{\m\n}(x)\right) \nonumber\\
 &=\int\!\frac{d^4k\, d^4x}{(2\pi)^4} \, \re ^{\ri k(y-x)}\star{\cal P}_\star\left(\int_0^1\!d\s F_{\m\a}(x+\s\th k)\, \th^{\a\b} (\ri k_\b) \star W(k,x)\star T^{\m\n}(x)\!\right)
 ,
\end{align}
where the star-commutator term arising from $\pa_\m T^{\m\n}$ was canceled by part of the contribution coming from $\pa_\m W$.
The factor $\ri k_\b$ can be pulled out of the integral by rewriting it as $\pa^y_\b$, thus allowing
for the definition of a gauge invariant, conserved (but no longer symmetric or traceless) \EMT $\bT^{\m\n}$:
\begin{align}
 \bT^{\m\n}& \equiv
 \tilde{T}^{\m\n}-\int\!\frac{d^4k\, d^4x}{(2\pi)^4}\, \re ^{\ri k(y-x)}\star{\cal P}_\star\left(\int_0^1\!d\s\, \th^{\m\a}F_{\a\b}(x+\s\th k)\star W(k,x)\star T^{\b\n}(x)\right)
 \,.
 \label{eq:gauge-em-tensor-final}
\end{align}
The fact that this modified \EMT is not traceless is actually not surprising since $\th^{\m\n}$
is not dimensionless and thereby
introduces a scale into the theory. However, sacrificing the
symmetry of the \EMT
will only be worth the price, if the construction above also works
when couplings to matter are considered.
In the following sections, we show that this is not the case.

\section{Coupling to neutral matter fields}\label{sec:matterfields1}

One of the peculiarities of non-commutative space is that even neutral matter (such as neutrinos) can couple to $U_\star (1)$ gauge fields (photons) via star-commutators~\cite{Hayakawa:1999,Chaichian:2001mu,Schupp:2002up},
i.e. the matter fields can transform with the adjoint representation (see \eqnref{eq:matteradjoint} below)
just like the gauge fields.
In the following, we study the \EMT of such neutral fields before discussing charged fields in \secref{sec:matterfields2}.

\subsection{Complex scalar field}

\paragraph{Scalar field action:\ }

We consider an external $U_\star(1)$ gauge field $(A^\m)$
and a complex scalar field $\phi$ in the \emph{adjoint representation,} i.e.
the infinitesimal gauge transformations read
\begin{align}
 \d_\l A_\m&=D_\m\l
 \, , \qquad
 \d_\l\phi=-\ig\starco{\phi}{\l}
 \, , \qquad
 \d_\l\phi^* =-\ig\starco{\phi^*}{\l}
 \,.
 \label{eq:matteradjoint}
\end{align}
The minimal coupling of the field $\phi$
to the external gauge field $(A^\m)$ is described by the action
\begin{align}
 S[\phi; A]=\inv{2}\intx\staraco{D_\m\phi^*}{D^\m\phi}
 \equiv \intx \, {\cal L}
 \,,\label{eq:action-phi}
\end{align}
where $D_\m\cdot=\pa_\m\cdot-\ig\starco{A_\m}{\cdot \, }$.
(In this respect we note that $D_\m \phi^* \equiv (D_\m \phi )^\dagger = \pa_\m \phi^* - \ig\starco{A_\m}{\phi^* }$.)
As has been argued in references~\cite{Hayakawa:1999,Chaichian:2001mu}, the fields in the adjoint
 representation carry zero $U_\star(1)$ charge. Concerning this point we note that
the derivative $D_\m\phi$ reduces in the commutative limit to $\pa_\m\phi$,
i.e. the coupling of neutral matter fields to gauge fields
is only possible in a non-commutative setting.

Due to the invariance of the integral under a cyclic permutation of the factors in the star-product,
the star-anticommutator in the action \eqref{eq:action-phi} has no effect,
but we choose to keep it in order to make manifest
the symmetry under the exchange $\phi\leftrightarrow\phi ^*$
in all expressions to be considered in the sequel.
The equations of motion for the scalar field read
\begin{align}
0=  \var{S[\phi;A]}{\phi^*}&=-D_\m\star D^\m\phi\,, &
0=  \var{S[\phi;A]}{\phi}&=-D_\m\star D^\m\phi^*\,,
\label{eq:EQMscal}
\end{align}
and we have
\begin{align}
J_\m \equiv 
 \var{S[\phi;A]}{A^\m}&=-\ig\big(
 \starco{\phi}{D_\m\phi^*}
+ (\phi \leftrightarrow \phi^* )
 \big)
 \,.
 \label{eq:matterJ}
\end{align}
This matter current (which vanishes in the commutative limit) is covariantly conserved, $ D^\m J_\m =0$,
as a consequence of the equations of motion~\eqref{eq:EQMscal} for $\phi$ and $\phi^*$.

The gauge transformation laws \eqref{eq:matteradjoint} imply that the covariant derivatives of $\phi$ and $\phi^*$ also transform covariantly, i.e.
 $\d_\l(D_\m\phi)=-\ig\starco{D_\m\phi}{\l}$
  and analogously for $D_\m\phi^*$.
  It follows that the Lagrangian density ${\cal L}$ in the action integral \eqref{eq:action-phi}
  transforms as $\d_\l {\cal L} =-\ig\starco{{\cal L}}{\l}$ so that the action is gauge invariant.
A short calculation using the Jacobi identity shows that the matter current~\eqref{eq:matterJ} also transforms covariantly, $\d_\l J_\m =-\ig\starco{J_\m}{\l}$.

Next we  turn to the \EMT of the model described by the action~\eqref{eq:action-phi}:
after coupling to an external gravitational field $\mg \equiv (g_{\m \n})$ we obtain the Einstein-Hilbert \EMT
in flat Moyal space:
\begin{align}
 T^{\m\n} &
 \equiv \left.\left( \frac{- 2}{\sqrt{|g|}}
 \var{S[\phi; A, \mg]}{g_{\m\n}}
 \right) \! \right|_{\mg=\id}
 =\inv2 \Big( \staraco{D^\m\phi^*}{D^\n\phi}
+ (\phi \leftrightarrow \phi^* )
 - \delta^{\m\n}
 \staraco{D_\r\phi^*}{D^\r\phi}  \Big)
 \,.
\label{eq:EMTNCscal}
\end{align}
From the equations of motion~\eqref{eq:EQMscal} it follows that
\begin{align}
 D_\m T^{\m\n}
 &=-\frac{\ig}{2}\Big(
 \staraco{D_\m\phi}{\starco{F^{\m\n}}{\phi^*}}
+ (\phi \leftrightarrow \phi^* )
 \Big)
 \,.
 \label{eq:DEMT}
\end{align}
 For non-Abelian Yang-Mills theory in commutative space there would be a trace on the right hand side
  and the cyclic invariance of this trace would enable us to rewrite it in terms of the
matter  current $J_\m$
 as ${\rm Tr}\, (F_{\n \m} J^\m )$
 (i.e. we have a continuum version of the non-Abelian Lorentz-force equation).
In the present case, however, all we can do is add and subtract the missing terms to arrive at
\begin{align}
 D_\m T^{\m\n}&=\inv2\staraco{F^{\m\n}}{J_\m}+\frac{\ig}{2}
 \Big(
 \starco{\phi}{\staraco{D_\m\phi^*}{F^{\m\n}}}
+ (\phi \leftrightarrow \phi^* )
 \Big)
 \,.
  \label{eq:DEMT2}
\end{align}
In commutative space, the second term would vanish under the trace.
In Moyal space, the cyclic  invariance
is only present under the integral $\intx$ (which in fact corresponds to a trace).
However, integrating the above equation is not very helpful, since
$\intx D_\m T^{\m\n}=0$
for the left hand side, i.e. it is a surface term and we would not get any new information.
In fact, as argued concerning equation~\eqref{eq:intlcl}, the second term on the right hand side of~\eqref{eq:DEMT2}
also vanishes upon integration with $\int d^3x$ in Minkowskian Moyal space with $\theta^{0i}=0$:
we will come back to this point after adding the gauge field action.

Another observation is that, just like the Lagrangian density,
the \EMT is not gauge invariant for the same reason, i.e.  lack of a trace
(hence of the cyclic invariance).
Instead, $T^{\m\n}$ transforms covariantly (as did its free gauge field counterpart discussed in the previous section):
\begin{align}
 \d_\l T^{\m\n} &=-\ig\starco{T^{\m\n}}{\l}
 \,.
 \label{eq:gtEMT}
\end{align}

\paragraph{Addition of the gauge field action:\ }

If we add to the action \eqref{eq:action-phi} a kinetic term
for the gauge field
so as to obtain the total action
\begin{align}
 S_{{\rm tot}} [\phi,A]=\inv{2}\intx\staraco{D_\m\phi^*}{D^\m\phi}
 + \frac{1}{4} \int d^4 x \, F^{\m\n} \star F_{\m\n}
 \,,\label{eq:action-phi+A}
\end{align}
then the equation of motion of $A_\m$ represents
the {\nc} version of Maxwell's  equations,
\begin{align}
 D^\m F_{\m\n}&=J_\n
 \,,
  \label{eq:F-eom}
\end{align}
where the expression of $J_\n$ in terms of $\phi$ and $\phi^*$ is given by equation~\eqref{eq:matterJ}.
From~\eqref{eq:F-eom}  and the argumentation in equation~\eqref{eq:covconserv}
it follows that the current $J^\m$ is covariantly conserved. Again, all fields and their covariant derivatives transform
covariantly under gauge variations.


The associated Einstein-Hilbert \EMT in flat Moyal space is a sum of expressions~\eqref{eq:EMTNCgauge} and~\eqref{eq:EMTNCscal}:
\begin{align}
 T^{\m\n}_{\textrm{tot}}&=\inv2\Big(
 \staraco{D^\m\phi}{D^\n\phi^*}
+ (\phi \leftrightarrow \phi^* )
 - \delta^{\m\n}
 \staraco{D_\r\phi^*}{D^\r\phi}\nn\\
 &\qquad\qquad \qquad\qquad \qquad\qquad
 +\staraco{F^{\m\r}}{F^{\n}_{\ \r}}-\inv2
 \delta^{\m \n}
 F_{\r\s}\star F^{\r\s}\Big)
 \,.
 \label{eq:EMTscalfield2}
\end{align}
Its covariant divergence can be obtained by using~\eqref{eq:F-eom} which implies,
for the gauge field EMT $T _{(A)} ^{\m\n}$,
\begin{align}
 D_\m T _{(A)} ^{\m\n}&= - \inv2\staraco{F^{\m\n}}{J_\m}
 \label{eq:div-A-emtensor}
\end{align}
and by adding~\eqref{eq:DEMT2} to this expression:
\begin{align}
 D_\m T^{\m\n}_{\textrm{tot}}&=\frac{\ig}{2}\big(
 \starco{\phi}{\staraco{D_\m\phi^*}{F^{\m\n}}}
+ (\phi \leftrightarrow \phi^* )
 \big)
 \,. \label{eq:div-total-emtensor}
\end{align}
In commutative space, there would be a trace on the right hand side so that this expression
would be zero and $T^{\m\n}$ would be conserved.
(In fact, in that case we would also have a trace on the r.h.s. of~\eqref{eq:EMTscalfield2},
and $ D_\m T^{\m\n}_{\textrm{tot}} =  \pa_\m T^{\m\n}_{\textrm{tot}}$.)
In the present case, however, we are once more lacking a trace to get rid of the r.h.s.
In this respect we emphasize that
according to the {\nc} generalization of Noether's theorem (see~\cite{Zahn:2003bt} and references therein),
a continuous symmetry of the action does not generally imply a standard local conservation law for interacting theories:
additional ``source'' terms (star-commutator terms which ultimately vanish under the space-time integral) generally appear.
Actually, integration of~\eqref{eq:div-total-emtensor}
with $\intx$ yields trivially zero on both sides (since integration corresponds to a trace).
In the Minkowskian version of Moyal space with non-commutativity parameters satisfying $\th^{0i}=0$,
it suffices to integrate over $\int\!d^3x$ to render the r.h.s. zero.
In this case we have
\begin{align}
0 =
 \int\! d^3x \, D_\m T^{\m\n}_{\textrm{tot}}
 =\int\! d^3x \, \pa_0 T^{0\n}_{\textrm{tot}}
 = \frac{d}{dt} \, \int\! d^3x \, T^{0\n}_{\textrm{tot}}
 \,,
 \label{eq:intltrafo}
\end{align}
which means that the four-momentum
$ P^\n \equiv  \int\!d^3x \, T^{0\n}_{\textrm{tot}}$
is a conserved quantity (which is also gauge invariant by virtue of~\eqref{eq:covtranslaw}, \eqref{eq:gtEMT}).
Of course the different contributions of
$ P^\n \equiv P_\phi^\n+P_A^\n$ are not conserved:
\begin{align}
 \pa_0 P_\phi^\n &=-\pa_0 P_A^\n =\int\!d^3x \,
 {F^{\n \m}} {J_\m (\phi)}
 \,.
\end{align}

\paragraph{Restoring gauge invariance:\ }

In the present setting,
we may follow the same strategy as in \secref{sec:gaugefields} and define an \EMT $\tilde T^{\m\n}$ which is gauge invariant
in analogy to expression~\eqref{eq:EMTGFM}:
\begin{align}
 \tilde T^{\m\n}_{\textrm{tot}}(y)&
 \equiv
 \int\!\frac{d^4k \, d^4x}{(2\pi)^4} \, \re^{\ri k(y-x)}\star W(k,x)\star T^{\m\n}_{\textrm{tot}}(x)
 \,. \label{eq:total-em-tensor-gauginv}
\end{align}
For its  divergence one obtains
\begin{align}
 \pa_\m^y \tilde T^{\m\n}_{\textrm{tot}}(y)&=\int\!\frac{d^4k \, d^4x}{(2\pi)^4}
\, \re^{\ri k(y-x)}\star\!\Bigg[{\cal P}_\star\left(\int_0^1\!d\s\, (\ri
k_\b) \th^{\b\a}F_{\a\m}(x\!+\!\s\th k)\star W(k,x)\star T^{\m\n}(x)\!\right)\nonumber\\
 &\quad\qquad +\frac{\ig}{2} W(k,x)\star \!\big(\!\starco{\phi^*}{\staraco{D_\m\phi}{F^{\m\n}}}+\starco{\phi}{\staraco{D_\m\phi^*}{F^{\m\n}}}\!\big)\Bigg]
 \,. \label{eq:cons-law-em-tens-gauginv}
\end{align}
The first term can be taken care of in the same way as in equation~\eqref{eq:gauge-em-tensor-final},
but not so the second term.
Thus, we are stuck with a small ($\th$-dependent)
breaking of $\pa_\m \tilde{T}^{\m\n}$, which of course vanishes in the commutative limit.

In Ref.~\cite{AbouZeid:2001up} a redefinition of the \EMT for a $\phi^{\star4}$ theory 
which implies its conservation was discussed.
However, in the present context the same strategy would destroy gauge covariance of $T^{\m\n}_{\textrm{tot}}$ making the construction of its gauge invariant
counterpart via Wilson line impossible, as we will now show.

Since the additional terms on the r.h.s. of \eqref{eq:div-total-emtensor} are star-commutators it is generally possible to pull out one derivative by making use of the so-called $\star'$-product (introduced in references~\cite{Liu:2000ad,Mehen:2000vs}),
\begin{align}
 (f\star'g)(x)
 \equiv
 \frac{\sin\left(\inv2{\pa}_\m^x\th^{\m\n}{\pa}_\n^y\right)}{\inv2{\pa}^x_\r\th^{\r\s}{\pa}^y_\s}f(x)g(y) \Big|_{x=y}
 \,, \label{eq:def-starprime}
\end{align}
so that we may write
\begin{align}
 \starco{f}{g}&=\ri\th^{\m\n}\pa_\m (f\star'\pa_\n g)
 \,.
\end{align}
Thus, a shift in the \EMT ensuing its conservation can be made in principle, 
but at the cost of destroying its gauge covariance, i.e.
\begin{align}
 \bar T^{\m\n}_{\textrm{tot}}&
 \equiv
 T^{\m\n}_{\textrm{tot}}+a\frac{g}{2}\th^{\m\s}\Big[\left(\phi^*\star'\pa_\s\staraco{D_\r\phi}{F^{\r\n}}+\phi\star'\pa_\s\staraco{D_\r\phi^*}{F^{\r\n}}\right)\nonumber\\
 &\quad\qquad -(1-a)\left(\pa_\s\phi^*\star'\staraco{D_\r\phi}{F^{\r\n}}+\pa_\s\phi\star'\staraco{D_\r\phi^*}{F^{\r\n}}\right)\Big]\,, \nonumber\\
 D_\m\bar T^{\m\n}_{\textrm{tot}}&=0\,, \qquad \bar T^{\m\n}_{\textrm{tot}}\neq \bar T^{\n\m}_{\textrm{tot}} \,,
 \qquad \d_\l\bar T^{\m\n}_{\textrm{tot}}\neq -\ig\starco{\bar T^{\m\n}_{\textrm{tot}}}{\l}
 \,,
\end{align}
where $a\in[0,1]$ is a free real parameter.
Similarly, a redefinition achieving $\pa_\m T^{\m\n}_{\textrm{tot}} =0$ could be made, but it would not gain us anything with respect to gauge invariance.

Thus, the best one can do is the gauge invariant expression \eqref{eq:total-em-tensor-gauginv} above with its modified conservation law \eqref{eq:cons-law-em-tens-gauginv}.
On the operator level, this means that only the trace over the divergence of the \EMT (which is an operator in quantized space) is conserved, a fact which is obscured by the star-product prescription where the trace becomes an integral over space(-time).
Therefore, it is not surprising that we have the local equation
$\pa_\m \tilde T^{\m\n}\neq0$, as was already observed in the case of the
$\phi^{\star4}$-theory in reference~\cite{Schweda:2000}. We note that
$\int\!d^4y\pa^y_\m \tilde T^{\m\n}_{\textrm{tot}}(y)=0=\intx\pa_\m^xT^{\m\n}_{\textrm{tot}}(x)$
as can be checked explicitly by
using the cyclic properties of the star-product under the integral, as well as $\int\!d^4k\exp(\ri ky)=(2\pi)^4\d^4(k)$ and $W(0,x)=1$.

In the next section, we show that one finds similar results
when coupling to fermions instead of (or in addition to) scalars.

\subsection{Fermions}
\label{sec:neutrinos}

We now consider an external $U_\star(1)$ gauge field
and a neutral fermionic (Dirac) field
(e.g. a {\nc} neutrino field~\cite{Schupp:2002up}), i.e. fields
in the \emph{adjoint representation} with the gauge transformation properties
\begin{align}
\d_\l\psi =-\ig\starco{\psi}{\l}
\,, \qquad
\d_\l\bpsi=-\ig\starco{\bpsi}{\l}
\,, \qquad
\d_\l A_\m =D_\m\l
\equiv
\pa_\m\l-\ig\starco{A_\m}{\l}
 \,. \label{eq:gauge-trafo-fermions}
\end{align}
The coupling of these fields is described by the action
\begin{align}
 S[\psi;A]&=\intx \, \ri \bpsi\star \g^\m D_\m\psi
 \, , \qquad {\rm where} \ \; D_\m\psi \equiv \pa_\m\psi-\ig \starco{A_\m}{\psi}
 \,,
 \label{eq:actDiracfield}
\end{align}
and $D_\m\bpsi \equiv \pa_\m\bpsi-\ig \starco{A_\m}{\bpsi \, }$, and
where the square
 matrices $\g^\m$ fulfill the Clifford algebra relation $\{\g^\m,\g^\n\}=2
 \delta^{\m \n} \id$.
The components
$\psi_{\alpha}$
of the spinor field $\psi$ are considered to be anticommuting (i.e. Grassmann) variables.
As for the scalar field case treated in the previous subsection, 
the coupling vanishes in the commutative limit where $D_\m\psi$ reduces to $\pa_\m\psi$.

The gauge transformation laws \eqref{eq:gauge-trafo-fermions} imply
\begin{align}
 \d_\l (D_\m\psi )&=-\ig\starco{D_\m\psi}{\l} \,, &
 \d_\l (D_\m\bpsi )&=-\ig\starco{D_\m\bpsi}{\l}
 \,,
\end{align}
so that the Lagrangian density $ {\cal L}$ of the model transforms as
$\d_\l {\cal L}  =-\ig\starco{ {\cal L}}{\l}$, hence the action is gauge invariant.

The equations of motion for $\psi$ and $\bpsi$ read
\begin{align}
 \g^\m D_\m\psi&=0= (D_\m\bpsi ) \g^\m
 \,,
\end{align}
and the fermionic
 matter current is given by
\begin{align}
 J^\m& \equiv \var{S}{A_\m}=-g\g^\m_{\a\b}\staraco{\psi_\b}{\bpsi_\a}
 \,,
\end{align}
where $\a,\b$ denote the spinor indices.
This current is covariantly conserved due to the equations of motion above (i.e.
$ D_\m J^\m=0$)
and it transforms covariantly under gauge transformations:
$ \d_\l J^\m=-\ig\starco{J^\m}{\l}$.
Therefore, we also have $\d_\l(D_\m J^\m)=-\ig[D_\m J^\m\stackrel{\star}{,}\l]=0$.
Furthermore, $J^\m$ vanishes in the commutative limit.

The action~\eqref{eq:actDiracfield} may also be written in the following more symmetric form
 by virtue of an integration by parts:
\begin{align}
 \intx \, \ri \bpsi\star \g^\m \overleftrightarrow{D}_\m\psi
 \equiv
 \intx \, \frac{\ri}{2} \Big( \bpsi\star \g^\m D_\m\psi- (D_\m\bpsi ) \star \g^\m \psi \Big)
 \, .
\end{align}
In analogy to the commutative case we obtain the \EMT for the Dirac
fields,\footnote{Its free counterpart (with coupling $g=0$) was previously constructed in reference~\cite{Ghasemkhani:2014zsa} 
where the Sugawara form of the \EMT for 
a free fermion in Moyal space was also established.}
\begin{align}
 T^{\m\n}&=\frac{\ri}{4}
 \Big[ \big( \bpsi\g^\m\star D^\n\psi
  - (D^\m\bpsi ) \g^\n\star\psi ) + ( \mu \leftrightarrow \nu \big)  \Big]
- \ri  \delta^{\m \n}
 \bpsi\star\g^\r \overleftrightarrow{D}_\r\psi \nn\\
 &=\frac{\ri}{4} \Big[ \big( \bpsi\g^\m\star D^\n\psi
 - (D^\m\bpsi ) \star\g^\n\psi  \big)
 +( \mu \leftrightarrow \nu ) \Big]
 \,,
\end{align}
where the second line holds on-shell due to the equations of motion.
Note that this tensor is hermitian as well as traceless on-shell.
Furthermore, it represents a sum of products of bilinear
quantities which transform covariantly, hence it also transforms covariantly
under the gauge transformations \eqref{eq:gauge-trafo-fermions},
\begin{align}
 \d_\l T^{\m\n}&=-\ig\starco{T^{\m\n}}{\l}
 \,,
\end{align}
as can be checked straightforwardly.

To evaluate the covariant divergence of the EMT, we note that the equation of motion
$\g^\n D_\n\psi=0$ implies
$0= (\g^\m D_\m) \star (\g^\n D_\n\psi) = \g^\m \g^\n D_\m D_\n \psi$:
by decomposing $\g^\m \g^\n$ into its 
symmetric and antisymmetric parts and by 
using the commutation relation
$\starco{D_\m}{D_\n} \psi = -\ri g  \starco{ F_{\m\n}}{\psi}$
it now follows that
\[
D^\m  D_\m  \psi = \frac{\ri}{2} \, g \g^\m \g^\n  \starco{ F_{\m\n}}{\psi}
\, .
\]
The algebra of $\g$-matrices then yields the result
\begin{align}
 D_\m T^{\m \n} & =
 g \bpsi \star \g_\m F^{\m \n} \star \psi
 - \frac{g}{2} \,  \staraco{F^{\m \n}}{\bpsi \star \g_\m \psi }
 + \frac{\ri}{4} \, \starco{\widetilde{F}^{\m\n}}{J^5_\m}
 \nn
 \\
 &=
 \frac12 \, \staraco{F^{\m\n}}{J_\m}
 +\frac{g}{2} \, \g^\m_{\a\b}
 \left( \staraco{\bpsi_\a}{F_\m^{\ \n} \star \psi_\b}+\staraco{\bpsi_\a \star F_\m^{\ \n}}{\psi_\b} \right)
 +\frac{\ri}{4} \, \starco{\widetilde{F}^{\m\n}}{J^5_\m}
 \,,
 \label{eq:EMTfermAdj}
\end{align}
where $\widetilde F^{\m\n} \equiv \frac12 \, \e^{\m\n\r\s}F_{\r\s}$
and $J^5_\m \equiv -g{\bpsi} \star \g^5\g_\m {\psi}$ denote the dual field strength tensor
and what may be interpreted as a chiral current, respectively.
The covariant divergence of $T^{\m \n}$
again involves additional terms which would vanish under a trace in Yang-Mills theory, resp. an integral in Moyal space
(the first term being compensated in the total \EMT  by the contribution~\eqref{eq:div-A-emtensor}
coming from the gauge field action).
It is interesting to note that similar additional terms were found in the context of matrix models\footnote{Note that in the matrix model context, scalars appear as extra dimensions and do not exhibit such extra terms due to internal symmetries~\cite{Okawa:2001if,Steinacker:2008ri}.} (cf. Appendix A.3  of reference~\cite{Polychronakos:2013fma}).

Once combined with the gauge field
EMT as discussed in the previous subsection, one may again define the
 gauge invariant counterpart
 of the total EMT   via Wilson line as in equation~\eqref{eq:total-em-tensor-gauginv}.
Once more $\pa_\m\tilde{T}^{\m\n}_{\textrm{tot}}$
involves  breaking terms which depend on the non-commutativity parameters $\theta^{\m \n}$ and
which cannot be absorbed into a redefined \EMT in a gauge invariant way.

\section{Coupling to charged matter fields}\label{sec:matterfields2}
In this section, we discuss the coupling of gauge fields to charged scalars and fermions.
In this case the \EMT is found to be gauge invariant in general so that one does not have to resort to  Wilson lines.
Nonetheless we will not have the standard local conservation law for this \EMT
and the latter cannot be achieved while maintaining  gauge invariance.

\subsection{Complex scalar field}

\subsubsection{Fundamental representation}
 A complex scalar field in Moyal space can also be coupled to a gauge field via the covariant derivative
 $\bD_\m \phi \equiv \pa_\m \phi -\ig A_\m \star \phi$ and its hermitian conjugate expression, i.e.
 \begin{align}
 \bD_\m\phi& \equiv \pa_\m\phi-\ig A_\m\star\phi\,, &
 \bD_\m^*\phi^*& \equiv (\bD_\m \phi )^\dagger = \pa_\m \phi^* + \ig \phi^* \star A_\m
 \,,
 \label{eq:covderivfundam}
\end{align}
  (where the latter derivative amounts to an action of $\pa_\m + \ig A_\m$ from the right).
Thus,  by contrast to the coupling discussed in the previous section,
the basic covariant derivatives presently do not involve a commutator of fields:
the field $\psi$ now transforms with the \emph{fundamental representation}, i.e.
we have the transformation rules
\begin{align}
 \d_\l\phi =\ig\l\star\phi\,, \qquad
 \d_\l\phi^* =-\ig\phi^*\star\l\,, \qquad
 \d_\l A_\m =D_\m\l \equiv \pa_\m\l-\ig\starco{A_\m}{\l}
 \,, \label{eq:scalar-gauge-trafo}
\end{align}
and thereby $\psi$ describes a charged field.

For a given external gauge field $A$, the action is now defined by
\begin{align}
 S[\phi; A]=\intx \, (\bD_\m^*\phi^*)( \bD^\m\phi)
 \,,
 \label{eq:action-phi2}
\end{align}
and the associated  equations of motion read
\begin{align}
0 =  \var{S[\phi;A]}{\phi^*}&=-\bD_\m\star \bD^\m\phi
\,, &
0=  \var{S[\phi;A]}{\phi}&=-\bD^*_\m\star \bD^{*\m}\phi^*
\,,
\end{align}
while
\begin{align}
J_\m \equiv 
 \var{S[\phi;A]}{A^\m} &= - \ig\big[
 \phi\star (\bD^*_\m\phi^* ) - ( \bD_\m\phi) \star\phi^*
 \big)]
 \,.
 \label{eq:scalar-eom}
\end{align}
The Lagrangian density ${\cal L} \equiv (\bD_\m^*\phi^*)\star ( \bD^\m\phi)$
in the action integral~\eqref{eq:action-phi2} is invariant under the gauge transformations \eqref{eq:scalar-gauge-trafo} which imply the transformation laws
\begin{align}
 \d_\l (\bD_\m\phi) =\ig\l\star \bD_\m\phi\,, \qquad
 \d_\l (\bD^*_\m\phi^*) =-\ig (\bD^*_\m\phi^*) \star\l
 \,. \label{eq:scalar-gauge-trafo2}
\end{align}
The matter current $J_\m$ transforms covariantly, $\d_\l J_\m =-\ig\starco{J_\m}{\l}$, and is covariantly conserved
by virtue of the equations of motion of $\phi$ and $\phi^*$:
\begin{align}
 D^\m J_\m& \equiv \pa^\m J_\m-\ig\starco{A^\m}{J_\m}=0
 \,. \label{eq:DJ-zero2}
\end{align}
The transformation property of $J_\m$ also leads to
$ \d_\l(D^\m J_\m)= -\ig\starco{D^\m J_\m}{\l}=0$
which  shows that the conservation law \eqref{eq:DJ-zero2} is gauge invariant as well, and thus provides a consistency check.

Before proceeding further, we note that the Lagrangian density is not uniquely defined.
For instance, instead of
${\cal L} \equiv (\bD_\m^*\phi^*)\star ( \bD^\m\phi)$ we can also consider
the same expression with a different order of factors,
${\cal L}' \equiv  ( \bD^\m\phi)\star
(\bD_\m^*\phi^*)$: both expressions differ by a total derivative, but ${\cal L}'$ is not gauge invariant,
$\delta_{\l} {\cal L}' = \ri g \starco{\l}{{\cal L}'}$.
A similar ambiguity occurs for the definition of the  corresponding \EMT which also represents a local, i.e.
non-integrated quantity.
As a matter of fact, we did not explicitly write any anticommutator in the action~\eqref{eq:action-phi2}
 in contrast to the scalar field model discussed in the previous section (where we had only one type of covariant derivative involving a commutator).
This is a choice that we may make (because the anticommutator can be dropped under the integral  in the action) and
the motivation for the present choice is that the Lagrangian density and the corresponding \EMT are now gauge invariant
(in contrast to our previous scalar field model):
the \EMT of the model~\eqref{eq:action-phi2} reads
\begin{align}
 T^{\m\n}& \equiv
 \left. \left(
 \frac{-2}{\sqrt{|g|}}\var{S[\phi;A, \mg ]}{g_{\m\n}} \right) \right|_{\mg=\id}
= (\bD^{*\,\m}\phi^* ) \star ( \bD^\n\phi )
 + ( \mu \leftrightarrow \nu )
 - \delta^{\m \n}
( \bD^*_\r\phi^* ) \star
( \bD^\r\phi )
 \, ,
 \label{eq:EMTASM}
\end{align}
and we have
\begin{align}
 \d_\l T^{\m\n}&=0
 \,,
 \label{eq:lgtEMT}
\end{align}
as follows readily  from the transformation laws\footnote{The \EMT $T'^{\m \n}$ corresponding to the Lagrangian
${\cal L}'  \equiv  ( \bD^\m\phi)\star (\bD_\m^*\phi^*)$,
i.e.
$T'^{\m\n}=( \bD^\m\phi ) \star (\bD^{*\,\n}\phi^* )+ ( \mu \leftrightarrow \nu )
 - \delta^{\m \n} ( \bD^\r\phi ) \star ( \bD^*_\r\phi^* )$
transforms covariantly under a gauge transformation (as does the one corresponding to the Lagrangian given by an anticommutator).
Furthermore, its covariant derivative produces additional commutator terms similar to the ones present on the r.h.s. of \eqnref{eq:fundamentalscalar-DT} below.} \eqref{eq:scalar-gauge-trafo2}.

Finally, using
\begin{align}
\starco{\bD_\m}{\bD_\n}\phi&=-\ig F_{\m\n}\star\phi\,, &
\starco{\bD^*_\m}{\bD^*_\n}\phi^*&=\ig\phi^*\star F_{\m\n}
\, ,
\end{align}
the divergence of the gauge invariant \EMT \eqref{eq:EMTASM} can be evaluated:
\begin{align}
 \pa_\m T^{\m\n}
 &=
 \ri g \left[ {\phi^*} \star F^{\m\n} \star (\bD_\m\phi) - ({\bD^*_\m \phi^*}) \star F^{\m\n} \star \phi \right]
 \nn
 \\
  &=F^{\m\n}\star J_\m +\ig\starco{\phi^*}{F^{\m\n}\star \bD_\m\phi}-\ig\starco{\bD^*_\m \phi^*}{F^{\m\n}\star\phi}
 \,.
 \label{eq:fundamentalscalar-DT}
\end{align}
Once more, we have non-vanishing commutator terms\footnote{Note that
$F^{\m\n}\star J_\m = \frac{1}{2} \, \staraco{F^{\m\n}}{J_\m} + \frac{1}{2} \, \starco{F^{\m\n}}{J_\m}$
where the first term is the opposite of the covariant divergence of the gauge field \EMT and where the second term is a 
star-commutator.}
due to a missing trace/integral (typical of {\nc} space)
even in this simpler  scalar field model.
By adding the gauge field action and integrating with $\int d^3x$ in Minkowskian Moyal space
with $\theta^{0i} =0$, we again find a conserved and gauge invariant four-momentum $(P^\n )$ by virtue of equations~\eqref{eq:intltrafo}
 and \eqref{eq:lgtEMT}.

Notice that, in contrast to the case of neutral fields, no Wilson line construction is necessary to define a gauge invariant EMT,
though in the present case the local conservation law of $T^{\m \n}$  is still broken by $\th$-dependent terms.
As mentioned in the previous section, in Ref.~\cite{AbouZeid:2001up} a trick was used to eliminate similar (commutator) terms from $\pa_\m T^{\m\n}$ in $\phi^{\star4}$-theory in Moyal space (but without coupling to a gauge field).
In that construction, one needs additionally the $\star'$-product \eqref{eq:def-starprime}.
We presently have
\begin{align}
 \starco{\phi^*}{F_{\m\n}\star \bD^\m\phi}=\ri\th^{\r\s}\pa_\r\left( \phi^*\star'\pa_\s \left(F_{\m\n}\star \bD^\m\phi\right)\right)=\ri\th^{\r\s}\pa_\s \left(\pa_\r \phi^*\star'\left(F_{\m\n}\star \bD^\m\phi\right)\right)
\end{align}
and the expression
\begin{align}
\bar{T}^{\m \n} _{\textrm{tot}} \equiv
 T_{\textrm{tot}}^{\m\n}&+ag\th^{\m\s}\left( \phi^*\star'\pa_\s \left(F^{\r\n}\star \bD_\r\phi\right)-\bD^{*}_\r\phi^*\star'\pa_\s\left( F^{\r\n}\star \phi\right)\right)\nonumber\\
 &-(1-a)g\th^{\m\s}\left( \pa_\s\phi^*\star' \left(F^{\r\n}\star \bD_\r\phi\right)-\pa_\s(\bD^{*}_\r\phi^*)\star'\left( F^{\r\n}\star \phi\right)\right)
 \,,
 \label{eq:primeEMT}
\end{align}
with $a\in[0,1]$ a free parameter, is a locally conserved quantity.
But the latter  is no longer gauge invariant, nor is it symmetric.

\subsubsection{Antifundamental representation}

We note that there exists another possible choice for the transformation laws of fields and for 
the related covariant derivatives~\cite{Hayakawa:1999,Chaichian:2001mu}.
This choice
is implemented by the replacement $\phi\leftrightarrow\phi^*$
(or the replacement $\bD_\m\leftrightarrow\bD^*_\m$ in integrands)
and amounts to assuming that the scalar field $\phi$ is in the
antifundamental representation rather than the fundamental representation (the latter corresponding to the transformation laws~\eqref{eq:scalar-gauge-trafo},
the covariant derivatives~\eqref{eq:covderivfundam} and the invariant action~\eqref{eq:action-phi2});
thus, we have the following expressions for the
\emph{antifundamental representation:}
 \begin{align}
 \d_\l\phi & =- \ig \phi \star \l \,, &
 \d_\l\phi^* &=\ig\l \star \phi^*
 \,,
 \nn\\
 \bD_\m^* \phi & \equiv \pa_\m\phi+\ig  \phi \star A_\m \,, &
 \bD_\m\phi^* & \equiv  \pa_\m \phi^* - \ig A_\m \star \phi^*
 \,,
 \nn\\
  S[\phi; A] & =\intx \, (\bD_\m^*\phi)( \bD^\m\phi ^*)
 \,.
\end{align}
Hence, we have a scalar field $\phi$ of opposite charge  (but the  additional change $g\to-g$
may  be considered to switch signs and thereby have the same charge for $\phi$ as in the fundamental representation).
This leads to the covariantly conserved current $\tilde J_\m=J_\m\big|_{\phi\leftrightarrow\phi^*}$
and similarly to the \EMT $\tilde{T}^{\m \n}$ which is related to the \EMT ${T}^{\m \n}$ of the fundamental representation
by the exchange $\phi\leftrightarrow\phi^*$.

\subsection{Fermions}

\subsubsection{Fundamental representation}

Finally, let us revisit
the coupling of fermions to an external gauge field by considering
the covariant derivative $\bar D_\m$:
\begin{align}
 S[\psi;A]&=\intx \, \ri \bpsi\star \g^\m \bD_\m\psi
\, , \qquad {\rm where} \ \;
 \bar D_\m \psi \equiv
 \pa_\m\psi-\ig A_\m\star\psi
 \,,
 \label{eq:simpDirac}
\end{align}
and $ \bD^*_\m\bpsi \equiv \pa_\m\bpsi+ \ri g\bpsi\star A_\m$.
Thus, we now consider fermions in the \emph{fundamental representation}~\cite{GraciaBondia:2000pz,Moreno:2000kt} (rather than the adjoint
as in \secref{sec:neutrinos}), the transformation laws being given by
\begin{align}
 \d_\l\psi =\ig\l\star\psi\,, \qquad
 \d_\l\bpsi =-\ig\bpsi\star\l \,, \qquad
 \d_\l A_\m =\pa_\m\l-\ig\starco{A_\m}{\l}
 \,. \label{eq:gauge-trafo-fermions2}
\end{align}
These transformations leave
the Lagrangian density ${\cal L} \equiv \ri \bpsi\star \g^\m \bD_\m\psi$
in the action functional~\eqref{eq:simpDirac}
invariant and they imply
\begin{align}
 \d_\l (\bD_\m\psi ) &=\ig\l\star \bD_\m\psi \,, &
 \d_\l ( \bD^*_\m\bpsi ) &=-\ig (\bD^*_\m\bpsi )\star\l
 \,.
\end{align}
The equations of motion of the present model read
\begin{align}
 \g^\m \bD_\m\psi&=0 \,, &
(\bD^*_\m\bpsi )  \g^\m =0
 \,, \label{eq:eomDiracsimp}
\end{align}
and the fermionic matter current is given by
\begin{align}
 J^\m& \equiv
 \var{S}{A_\m}=-g\g^\m_{\a\b}\psi_\b\star\bpsi_\a
 \,. \label{eq:chargedfermionJ}
\end{align}
It is covariantly conserved by virtue of the equations of motion~\eqref{eq:eomDiracsimp}, i.e.
$D_\m J^\m \equiv \pa_\m J^\m-\ig\starco{A_\m}{J^\m}=0$.
Furthermore, it transforms covariantly under gauge transformations,
$ \d_\l J^\m=-\ig\starco{J^\m}{\l}$.
Thus, we also have $\d_\l(D_\m J^\m)=0$.

The (on-shell) expression for the  \EMT of fermion fields reads
\begin{align}
 T^{\m\n} &=\frac{\ri}{4}
 \left[ \left( \bpsi\g^\m\star \bD^\n\psi -(\bD^{*\m}\bpsi) \star\g^\n\psi \right) + (\mu \leftrightarrow \nu ) \right]
 \,. \label{eq:chargedfermionEMT1}
\end{align}
It is traceless on-shell and (just like the Lagrangian density) it is invariant under the gauge transformations~\eqref{eq:gauge-trafo-fermions2}.
Thus, no Wilson line construction is needed here either.

The divergence of this \EMT can be determined by using the equations of motion:
\begin{align}
 \pa_\m T^{\m\n} &= g \bpsi\g_\m\star F^{\m\n}\star\psi
 =  F^{\m\n}\star J_\m + g
  \g^\m_{\a\b}\staraco{\bpsi_\a}{F_\m^{\ \n}\star\psi_\b}
 \,.
\end{align}
Once again, there is an additional term which would vanish under a trace, resp. an integral.
Since  the  fermions are Grassmann variables, the additional term may be written  in terms of a
 $\star'$-product as
\begin{align}
 \staraco{\bpsi_\a}{F_\m^{\ \n}\star\psi_\b}&=i\th^{\r\s}\pa_\r\bpsi_\a\star'\pa_\s(F_\m^{\ \n}\star\psi_\b)
 \,.
\end{align}
This allows for a redefinition of $T^{\m\n}_{\textrm{tot}}$ which is conserved but not gauge invariant.

Before closing this section, let us see what happens if we consider the \EMT
corresponding to the classically equivalent Lagrangian ${\cal L}' \equiv - \ri  \g^\m_{\a\b} (\bD_\m\psi_{\b}) \star \bpsi_{\a}$:
\begin{align}
 T'^{\m\n} &=\frac{-\ri}{4}
 \left[ \left(\g^\m_{\a\b} ( \bD^\n\psi_\b ) \star \bpsi_\a -  \g^\m_{\a\b}\psi_\b \star(\bD^{*\n} \bpsi_\a) \right) + (\mu \leftrightarrow \nu ) \right]
  \, .
\label{eq:altEMT}
\end{align}
This expression is not gauge invariant, but transforms covariantly: $\d_\l T'^{\m\n} =-\ig\starco{T'^{\m\n}}{\l}$.
For the covariant divergence of the EMT~\eqref{eq:altEMT}  we get a result which bears some resemblance
with the expression~\eqref{eq:EMTfermAdj} obtained for fermion fields in the adjoint representation,
\begin{align}
 D_\m T'^{\m\n}&= \frac12 \staraco{F^{\m\n}}{J_\m}
 - \frac{\ri}{4} \, \starco{\widetilde{F}^{\m\n}}{J^5_\m}
 \,,
\end{align}
where $\widetilde F^{\m\n} \equiv \frac12 \, \e^{\m\n\r\s}F_{\r\s}$
and $J^5_\m \equiv -g (\g^5\g_\m)_{\a\b}\, {\psi_\b}\star {\bpsi_\a}$.
Thus, we again have an additional commutator term in the covariant
divergence of the EMT.

\subsubsection{Antifundamental representation}

Roughly speaking, the interchange of $\psi_{\alpha}$ and $\bpsi_{\beta}$
(or of  $\bD$ and $\bD^*$ in integrands) in the previous expressions allows us to obtain the fermion field  (of opposite charge)
in the \emph{antifundamental representation:}
\begin{align}
\d_\l\psi & = -\ig\psi\star\l
\,, &
 \d_\l\bpsi &= \ig \l\star\bpsi
\,,
 \nn\\
 \bD^*_\m\psi & \equiv \pa_\m \psi
 + \ri g\psi\star A_\m
\,, &
\bD_\m \bpsi &\equiv
 \pa_\m\bpsi-\ig A_\m\star\bpsi
 \,,
 \nn\\
  S[\psi;A]&=\intx \, \ri \bpsi\star \g^\m \bD^*_\m\psi
 \, .
\end{align}
In this case, the covariantly conserved current
takes the familiar form
\begin{align}
\tilde{J}^\m=-g \bpsi  \star \g^\m \psi
\, ,
\end{align}
i.e. \eqnref{eq:chargedfermionJ} with  $\bpsi_{\alpha}$ and $\psi_{\beta}$
exchanged.
The resulting expression
 for the \EMT in the anti\-fundamental representation is obtained by exchanging  $\bD_\m$ and $\bD^*_\m$
in expression \eqref{eq:chargedfermionEMT1}:
 $\tilde{T}^{\m \n} = \frac{\ri}{4}\!\!
 \left[ \left( \bpsi\g^\m\star \bD^{*\n}\psi -(\bD^{\m}\bpsi) \star\g^\n\psi \right) + (\mu \leftrightarrow \nu ) \right]$.
 Just like the Lagrangian density $\tilde{\cal L} \equiv \ri \bpsi \star \g^\m \bD^{*}_{\m}\psi$, this tensor transforms covariantly with the
 adjoint representation under gauge transformations.
For the covariant divergence of the EMT we have results which are similar to those holding in the fundamental representation.

\section{Conclusion}
According to the {\nc} generalization of Noether's theorem~\cite{Zahn:2003bt},
some extra $\th$-dependent terms (``source''/star-commutator terms)
generally appear in the local conservation law for the EMT for interacting theories.
The lack of gauge invariance and local conservation of the \EMT is not surprising since the EMT represents, very much like the Lagrangian density,
a non-integrated expression and it is only the integral over Moyal space
which ensures the cyclic invariance of factors in star-products,
and thereby the vanishing of star-commutator terms.

In the present paper, we have explicitly
shown (for complex scalars as well as for fermions coupled to gauge fields)
that the standard local conservation law of the \EMT $T^{\m\n}$ is always modified due to {\nc} effects and that
$T^{\m\n}$ can always be redefined so as to be conserved, but that the so defined EMT is not gauge invariant.
(Yet, for dynamical matter and gauge fields we always have a conserved and gauge invariant four-momentum with components
$P^{\nu} = \int d^3x \, T^{0\nu}$.)

More specifically, we discussed two possible couplings of scalars and fermions to gauge fields corresponding to neutral and charged matter, respectively:
In the first case, the basic \EMT transforms covariantly and its gauge invariant counterpart could be constructed by using the {\nc} generalization of a Wilson line.
In the second case (for which there exist two variants, namely the fundamental and the antifundamental representations),
the freedom in the definition of the \EMT allows
for the choice of a gauge covariant or a gauge invariant tensor.
For all cases we found that the consideration of the $\star'$-product allows to achieve
the standard local conservation law for the EMT, but at the expense of losing gauge invariance (and symmetry).
We note that the tools employed here are also those which are generally
considered for the quantization, e.g. see references~\cite{Gross:2000ba, Liu:2000ad, Bellucci:2003ud}.

Our systematic study is tantamount to a proof that it is not possible to construct a conserved and gauge invariant
(and symmetric) EMT for spin $0$ and spin $1/2$ matter fields coupled to a $U_{\star}(1)$ gauge field in Moyal
space\footnote{An indirect cure of the problems  for the case of neutral scalars appears to be the passage to the matrix model framework
since these scalars
appear naturally as extra dimensions in this framework
and the extra terms we found in the conservation law of the EMT are not present there 
due to internal symmetries~\cite{Okawa:2001if,Steinacker:2008ri}.}.
Yet, in all cases the total energy-momentum $P^\n \equiv \int d^3x \, T^{0\n}_{{\rm tot}}$
of the system represents a conserved and gauge invariant quantity.
In practice, the formulation of classical as well as quantum field theories in flat space primarily relies
on the conserved charges $P^\n$
so that the problematic properties of the EMT that we discussed can somehow be circumvented.
However, the situation is quite different in curved space where one has to couple the \EMT to a metric field
while taking into account the related non-commutativities.

\subsection*{Acknowledgements}
D.~B. wishes to thank H.~Steinacker for pointing out to us Ref.~\cite{Polychronakos:2013fma}.
Furthermore, we wish to thank the anonymous referees for their valuable comments
which helped us to clarify several points.


\providecommand{\href}[2]{#2}\begingroup\raggedright\endgroup

\end{document}